\documentclass{article}

\newcommand{\beq}{\begin{equation}}
\newcommand{\eeq}{\end{equation}}
\newcommand{\ket}[1]{\left| {#1} \right>}
\newcommand{\bra}[1]{\left< {#1} \right|}
\newcommand{\braket}[2]{\left< {#1} \right. \left| {#2} \right>}

\newcommand{\subsec}[1]{\vspace{12pt}\noindent {\em {#1}}\vspace{12pt}}

\usepackage{graphics}

\begin{document}

\title{Quantum Computing and Error Correction}
\author{Andrew M. Steane\\
\em Centre for Quantum Computation,\\
\em Department of Physics, University of Oxford,\\
\em Clarendon Laboratory, Parks Road, Oxford OX1 3PU, England.
}

\date{November 8, 2000}

\maketitle

\begin{abstract}
The main ideas of quantum error correction are introduced. These
are encoding, extraction of syndromes, error operators, and code
construction. It is shown that general noise and relaxation of a
set of 2-state quantum systems can always be understood as a
combination of Pauli operators acting on the system. Each quantum
error correcting code allows a subset of these errors to be
corrected. In many situations the noise is such that the remaining
uncorrectable errors are unlikely to arise, and hence quantum
error correction has a high probability of success. In order to
achieve the best noise tolerance in the presence of noise and
imprecision throughout the computer, a hierarchical construction
of a quantum computer is proposed.
\end{abstract}

\maketitle

\section{Introduction}

Quantum error correction (QEC) is one of the basic components of
quantum information theory. It arises from the union of quantum
mechanics and classical information theory, and leads to some
powerful and surprising possibilities in the quantum context. For
example, we may use a quantum error correcting code to allow us to
store a general, unknown state of a two-state system in a set of
five two-state systems (eg two-level atoms), in such a way that if
any unknown change (arbitrarily large and including the case of
relaxation) should occur in the state of any one atom, then it is
still possible to recover the originally stored state precisely:
the seemingly large perturbation had no effect at all on the
stored information.

QEC is by now sufficiently well established that it has chapters
devoted to it in text books, for example
\cite{Bk:Lo,Bk:Bouwmeester,Bk:Nielsen}. Nevertheless, an
introduction to the ideas is still useful for people new to the
subject, so much of the discussion here will be devoted to this.
I will then discuss recent research in section 7.

QEC may be considered to consist of two separate parts. The first
part is the theory of quantum error correcting codes, syndromes
and quantum error operators. This is essentially a collection of
physical insights and mathematical ideas which follow directly
from the principles of quantum mechanics and are uncontroversial.
The second part is the physics of noise, which must include the
theory of how an open quantum system interacts with its
environment. The success of QEC methods in practice will depend on
our having the correct understanding of noise in open quantum
systems. There is still some controversy over which methods can be
correctly applied when we are considering large-scale systems such
as a `classical' environment, but the success of QEC is found not
to depend on special pleading in this area. Nevertheless, it is
necessary to be careful to examine what assumptions are required,
so the presentation here will focus on this, particularly
in sections 4 and 6.

\section{Background}

The first quantum error correcting codes were discovered independently by
Shor \cite{95:Shor} and Steane \cite{96:SteaneA}. Shor proved that 9 qubits could
be used to protect a single qubit against general errors, while Steane
described a general code construction whose simplest example does the same
job using 7 qubits (see section 5.1). A general theory of quantum
error correction dates from subsequent papers of Calderbank and Shor \cite
{96:Calderbank} and Steane \cite{96:SteaneB} in which general code constructions,
existence proofs, and correction methods were given. Knill and Laflamme \cite
{97:Knill} and Bennett {\em et. al.} \cite{96:Bennett}
provided a more general theoretical framework, describing
requirements for quantum error correcting codes, and
measures of the fidelity of corrected states.

The important concept of the stabilizer (section 5.1) is due to
Gottesman \cite{96:Gottesman} and independently Calderbank {\em
et. al.} \cite{97:Calderbank}; this yielded many useful insights
into the subject, and permitted many new codes to be discovered
\cite{96:Gottesman,97:Calderbank,99:SteaneC}. Stabilizer methods
will probably make a valuable contribution to other areas in
quantum information physics. The idea of recursively encoding and
encoding again was explored by several authors
\cite{96:Knill,98:Aharonov,97:KitaevA}, this uses more quantum
resources in a hierarchical way, to permit communication over
arbitrarily long times or distances. Van Enk {\em et. al.}
\cite{97:vanEnkA,97:vanEnkB} have discussed quantum communication
over noisy channels using a realistic model of trapped atoms and
high-quality optical cavities, and recursive techniques for
systems in which two-way classical communication is possible were
described by Briegel {\em et al.} \cite{98:Briegel}.

Quantum error correction is discussed in reviews of quantum
information theory, see for example \cite{98:Steane,98:Bennett}.

\section{Three bit code}
\label{s:3bit}

We will begin by analysing in detail the workings of the most simple quantum
error correcting code. Exactly what is meant by a quantum error correcting
code will become apparent.

Suppose a source {\sc A} wishes transmit quantum information via a noisy
communication channel to a receiver {\sc B}. Obviously the channel must be
noisy in practice since no channel is perfectly noise-free. However, in
order to do better than merely sending quantum bits down the channel, we
must know something about the noise. For this introductory
section, the following properties will be assumed: the noise acts on each
qubit independently, and for a given qubit has an effect chosen at random
between leaving the qubit's state unchanged (probability $1-p$) and applying
a Pauli $\sigma_x$ operator (probability $p < 1/2$). This is a very artificial
type of noise, but once we can correct it, we will find that our correction
can also offer useful results for much more realistic types of noise.

\begin{figure}[ht]
\centerline{\resizebox{!}{5 cm}{\includegraphics{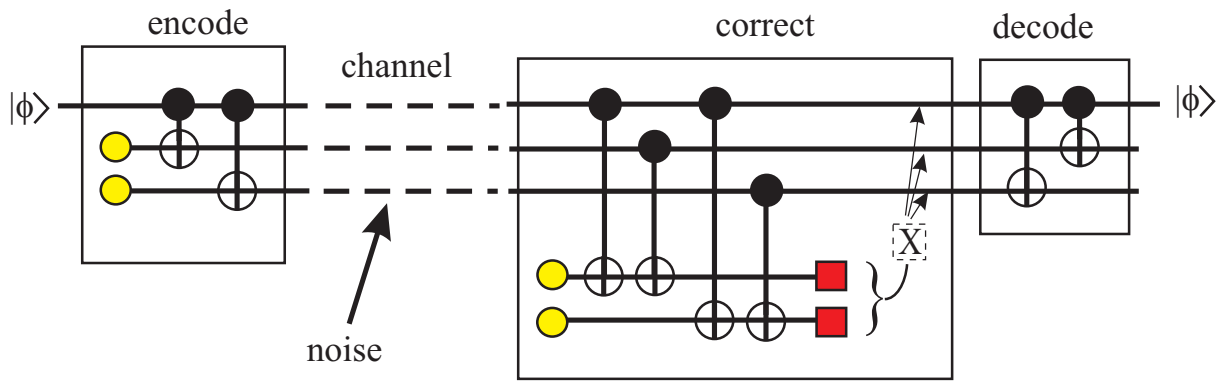}}}
%\rule{0pt}{24 pt}
\caption{Example of quantum error correction which illustrates the main features.
A single qubit is encoded into three, and these are sent down the noisy channel.
The receiver introduces further qubits (the ancilla) and uses them to extract a
syndrome by applying controlled-not gates and then measuring the ancilla bits. The
received state can then be corrected. Finally, the single qubit of information may be
re-extracted (decoded) from the three.}
\end{figure}

The simplest quantum error correction method is summarised in fig. 1. We
adopt the convention of calling the source Alice and the receiver Bob. The
state of any qubit which Alice wishes to transmit can be written without
loss of generality $a\left| {0} \right> + b\left| {1} \right>$. Alice
prepares two further qubits in the state $\left| {0} \right>$, so the
initial state of all three is $a\left| {000} \right> + b \left| {100} \right>
$. Alice now operates a controlled-{\sc not} gate from the first qubit to
the second, producing $a\left| {000} \right> + b \left| {110} \right>$,
followed by a controlled-{\sc not} gate from the first qubit to the third,
producing $a\left| {000} \right> + b \left| {111} \right>$. Finally, Alice
sends all three qubits down the channel.

The states $\ket{000}$ and $\ket{111}$ are called quantum
codewords. Only codewords, or superpositions of them, are sent by
Alice.

Bob receives the three qubits, but they have been acted on by the noise in
the channel. Their state is one of the following:
\begin{eqnarray}
\begin{array}{ll}
\mbox{state} & \mbox{probability} \\
a\left| {000} \right> + b \left| {111} \right> & (1-p)^3 \\
a\left| {100} \right> + b \left| {011} \right> & p(1-p)^2 \\
a\left| {010} \right> + b \left| {101} \right> & p(1-p)^2 \\
a\left| {001} \right> + b \left| {110} \right> & p(1-p)^2 \\
a\left| {110} \right> + b \left| {001} \right> & p^2(1-p) \\
a\left| {101} \right> + b \left| {010} \right> & p^2(1-p) \\
a\left| {011} \right> + b \left| {100} \right> & p^2(1-p) \\
a\left| {111} \right> + b \left| {000} \right> & p^3
\end{array}
\end{eqnarray}
Bob now introduces two more qubits of his own, prepared in the state $\left|
{00} \right>$. This extra pair of qubits, referred to as an {\em ancilla},
is not strictly necessary, but makes error correction easier to understand
and becomes necessary when fault-tolerant methods are needed. Bob uses the
ancilla to gather information about the noise. He first carries out
controlled-{\sc not}s from the first and second received qubits to the first
ancilla qubit, then from the first and third received qubits to the second
ancilla bit. The total state of all five qubits is now
\begin{eqnarray}
\begin{array}{cc}
\mbox{state} & \mbox{probability} \\
(a\left| {000} \right> + b \left| {111} \right>)\left| {00} \right> & (1-p)^3
\\
(a\left| {100} \right> + b \left| {011} \right>)\left| {11} \right> &
p(1-p)^2 \\
(a\left| {010} \right> + b \left| {101} \right>)\left| {10} \right> &
p(1-p)^2 \\
(a\left| {001} \right> + b \left| {110} \right>)\left| {01} \right> &
p(1-p)^2 \\
(a\left| {110} \right> + b \left| {001} \right>)\left| {01} \right> &
p^2(1-p) \\
(a\left| {101} \right> + b \left| {010} \right>)\left| {10} \right> &
p^2(1-p) \\
(a\left| {011} \right> + b \left| {100} \right>)\left| {11} \right> &
p^2(1-p) \\
(a\left| {111} \right> + b \left| {000} \right>)\left| {00} \right> & p^3
\end{array}
\label{sstate}
\end{eqnarray}
Bob measures the two ancilla bits in the basis $\{\left| {0} \right>, \left|
{1} \right>\}$. This gives him two classical bits of information. This
information is called the {\em error syndrome}, since it helps to diagnose
the errors in the received qubits. Bob's next action is as follows:
\[ \mbox{\begin{tabular}{cl}
measured syndrome & \hspace{3em} action \\
00 & do nothing \\
01 & apply $\sigma_x$ to third qubit \\
10 & apply $\sigma_x$ to second qubit \\
11 & apply $\sigma_x$ to first qubit
\end{tabular}}
\]
Suppose for example that Bob's measurements give $10$ (i.e. the ancilla
state is projected onto $\left| {10} \right>$). Examining eq. (\ref{sstate}%
), we see that the state of the received qubits must be either $a\left| {010}
\right> + b \left| {101} \right>$ (probability $p(1-p)^2$) or $a\left| {101}
\right> + b \left| {010} \right>$ (probability $p^2(1-p)$). Since the former
is more likely, Bob corrects the state by applying a Pauli $\sigma_x$
operator to the second qubit. He thus obtains either $a\left| {000} \right>
+ b \left| {111} \right>$ (most likely) or $a\left| {111} \right> + b \left|
{000} \right>$. Finally, to extract the qubit which Alice sent, Bob applies
controlled-{\sc not} from the first qubit to the second and third, obtaining
either $(a\left| {0} \right> + b\left| {1} \right>)\left| {00} \right>$ or $%
(a\left| {1} \right> + b \left| {0} \right>)\left| {00} \right>$. Therefore
Bob has either the exact qubit sent by Alice, or Alice's qubit operated on
by $\sigma_x$. Bob does not know which he has, but the important point is
that the method has a probability of success greater than $1-p$. The
correction is designed to succeed whenever either no or just one qubit is
corrupted by the channel, which are the most likely possibilities. The
failure probability is the probability that at least two qubits are
corrupted by the channel, which is $3p^2(1-p) + p^3 = 3p^2 - 2p^3$, i.e.
less than $p$ (as long as $p<1/2$).

To summarise, Alice communicates a single general qubit by expressing its
state as a joint state of three qubits, which are then sent to Bob. Bob
first applies error correction, then extracts a single qubit state. The
probability that he fails to obtain Alice's original state is $O(p^2)$,
whereas it would have been $O(p)$ if no error correction method had been
used. We will see later that with more qubits the same basic ideas lead to
much more powerful noise suppression, but it is worth noting that we already
have quite an impressive result: by using just three times as many qubits,
we reduce the error probability by a factor $\sim 1/3p$, i.e. a factor $%
\sim\! 30$ for $p=0.01$, $\sim\! 300$ for $p=0.001$, and so on.

\newpage

\subsec{3.1 Phase errors}

The channel we just considered was rather artificial. However, it
is closely related to a more realistic type of channel: one which
generates random rotations of the qubits about the $z$ axis. Such
a rotation is given by the operator
\begin{equation}
P(\epsilon \phi) = \left( \!\!\!\! \begin{array}{cc}
e^{i \epsilon \phi} & 0 \\
0 & e^{-i \epsilon \phi} \end{array} \!\!\!\! \right) \equiv  \cos( \epsilon \phi ) I +
i \sin(\epsilon \phi) \sigma_z           \label{P}
\end{equation}
where $I$ is the identity, $\epsilon$ is a fixed quantity indicating the typical
size of the rotations, and $0 < \phi < 2 \pi$ is a random angle. We may
understand such an error as a combination of no error ($I$) and a
phase flip error ($\sigma_z$). We can deal with this situation by
using exactly the same quantum error correcting code as before,
only now at either end of the channel we apply to each qubit the Hadamard
rotation
\[
H \equiv \frac{1}{\sqrt{2}} \left( \begin{array}{cc} 1 & 1 \\ 1 &
-1 \end{array} \right)
\]
The combined effect of phase error and these Hadamard rotations is
\beq
H P H = \cos( \epsilon \phi ) I + i \sin(\epsilon \phi) \sigma_x
\eeq
which is easy to derive using $HH = I$ and $H \sigma_z H = \sigma_x$.

We now have a situation just like that considered in the previous
section, only instead of a $\sigma_x$ `bit flip' being applied
randomly with probability $p$ to each qubit, every qubit certainly
experiences an error which is a combination of the identity and
$\sigma_x$. When we put this type of error through the analysis of
the correction network shown in figure 1, modeling the
measurements on the ancilla qubits as standard Von-Neumann  projective
measurements, the outcome is exactly as before, with the quantity $p$
equal to the average weight of the terms in the quantum
state which have single bit-flip errors (before correction). This average is over the
random variable $\phi$, thus $p = \left< \sin^2 \epsilon \phi \right> \simeq
(2 \pi \epsilon)^2/3$ for $\epsilon \ll 1$, where
we have considered the worst case, in which the states being transmitted
are taken to orthogonal states by the action of $\sigma_z$.
A complete
calculation of the 3-qubit system, with the random rotations and the
syndrome measurement described, confirms that the fidelity of the final
corrected state is $f = 1 - 3 p^2$ for small $p$, in agreement
with the results of the previous section.

\subsec{3.2 Projective errors}

It is a familiar feature of quantum mechanics that a set of
particles where each is in an equal superposition of two states
$(\ket{0} + \exp(i \phi) \ket{1})/\sqrt{2}$, with the relative
phase $\phi$ of the two terms random, is indistinguishable from a statistical
mixture, i.e. a set of particles where each is either in the state $\ket{0}$ or in
$\ket{1}$, randomly. This follows immediately from the fact that
these two cases have the same density matrix. With this in mind, it
should not be surprising that another type of error which the
three-bit code can correct is an error consisting of a projection
of the qubit onto the $\ket{0}, \ket{1}$ basis. To be specific,
imagine the channel acts as follows: for each qubit passing
through, either no change occurs (probability $1-2p$) or a
projection onto the $\ket{0}, \ket{1}$ basis occurs (probability
$2p$). The projection $\ket{0}\bra{0} = (I + \sigma_z)/2$ and
$\ket{1}\bra{1} = (I - \sigma_z)/2$. Such errors are identical
to phase errors (equation(\ref{P})), except for the absence of
the factor $i$ before the $\sigma_z$ term. However, this factor
does not affect the argument, and once again the analysis in
equation (\ref{sstate}) applies (once we have used the Hadamard
`trick' to convert phase errors to bit-flip errors) in the limit
of small $p$. Note that we define $p$ in terms of the effect of
the noise on the state: it is not the probability that a
projection occurs, but the probability that a
$\sigma_z$ error is produced in the state of any single qubit
when quantum codewords are sent down the channel.

Another way of modeling projective errors is to consider that
the projected qubit is first coupled to some other system, and
then we ignore the state of the other system. Let such another
system have, among its possible states, two states
$\ket{\alpha}_e$ and $\ket{\beta}_e$ which are close to one
another, $\braket{\alpha}{\beta} = 1 - \epsilon$.
The error consists in the
following coupling between qubit and extra system:
\beq
\left( a \ket{0} + b \ket{1} \right) \ket{\alpha}_e  \rightarrow
a \ket{0} \ket{\alpha}_e + b \ket{1} \ket{\beta}_e
\eeq
this is an entanglement between the qubit and the extra system. A
useful insight is to express the entangled state in the following
way:
\beq
a \ket{0} \ket{\alpha} + b \ket{1} \ket{\beta} \equiv
 \frac{1}{\sqrt{2}} \left[ \left( a \ket{0} + b \ket{1} \right)
\ket{+}_e  + \left( a \ket{0} - b \ket{1} \right)
\ket{-}_e \right]
\eeq
where $\ket{\pm}_e = (\ket{\alpha}_e \pm \ket{\beta}_e)/\sqrt{2}$.
Hence the error on the qubit is seen to be a combination of
identity and $\sigma_z$, and is correctable as before. The
probability that the $\sigma_z$ error is produced is calculated
by finding the weights of the different possibilities in equation
(\ref{sstate}) after the tracing over the extra system. We thus
obtain $p = ( 1- {\rm Re}(\braket{\alpha}{\beta}))^2 / 2
= \epsilon^2 / 2$.

\section{Most general possible error, digitization of noise}

The most general possible error that a qubit can undergo is a
general combination of the ones we have listed, that is, general
rotations in the two-dimensional Hilbert space, combined with a
possible projection onto any axis. In what follows we will model
a projection by using the method just outlined, in which the
qubit is entangled with another system, and the error consists in
the fact that the entangling process can not be undone, because
the other system is not under our control. In QEC it has become
standard practice to refer to coupling to `the environment' in
this way. It might seem that this depends on modeling the
environment as if it is a quantum system with unitary evolution,
but in fact the technique is only a mathematical convenience for
dealing with projections. If one is uncomfortable with writing
down quantum states of the environment, then one should return to
the language of projection, and the conclusions concerning the
efficacy of QEC will remain unchanged.

With these considerations in mind, we now come to one of the basic
insights of QEC: a completely arbitrary change in a general state
$\ket{\phi}$ of a qubit can be written
  \beq
\ket{\phi} \ket{\psi_0}_e \rightarrow \sum_{i=\{I,x,y,z\} } \left(
\sigma_i \ket{\phi} \right) \ket{\psi_i}_e    \label{1error}
  \eeq
where we have written $I = \sigma_I$ and the second system with
states $\ket{\psi_i}_e$ (typically called `states of the
environment') is introduced simply to allow us to write using the
language of kets a general change in the density matrix of the
qubit. This density matrix of the qubit is the reduced density
matrix of the pair of systems, after tracing over the second
system. We can thus reproduce arbitrary changes in the density
matrix of the qubit as long as we allow the states
$\ket{\psi_i}_e$ to be unconstrained, that is, they are not
necessarily either normalised or orthogonal.

Generalising this insight, we find that any transformation of the state
of a set of qubits can be expressed
\beq
\ket{\phi} \ket{\psi_0}_e \rightarrow \sum_i
\left(E_i \ket{\phi} \right) \ket{\psi_i}_e   \label{noise}
\eeq
where each `error operator' $E_i$ is a tensor product of Pauli
operators acting on the qubits, $\ket{\phi}$ is the initial state of
the qubits, and $\ket{\psi_i}_e$ are states of a second system, not
necessarily orthogonal or normalised.

We thus express general noise and/or decoherence in terms of
Pauli operators $\sigma_x$, $\sigma_y$, $\sigma_z$ acting on
the qubits. These will be written $X\equiv \sigma_x$,
$Z \equiv \sigma_z$, $Y \equiv -i\sigma_y = X Z$. The remarkable
fact embodied by (\ref{noise}) is that arbitrary changes, which
form a continuous set, can be `digitized' by expressing them as
a weighted sum of discrete errors. This arises from the fact that
the Pauli matrices with the identity form a complete set.

To write tensor products of Pauli matrices acting on $n$ qubits,
we introduce the notation $X_u Z_v$ where $u$ and $v$ are $n$-bit
binary vectors. The non-zero coordinates of $u$ and $v$ indicate
where $X$ and $Z$ operators appear in the product. For example,
  \beq X \otimes I \otimes Z \otimes Y \otimes X \equiv X_{10011}
Z_{00110}.
  \eeq
The {\em weight} of an error operator is the number of terms in
the tensor product which are not the identity, this is 4 in the
example just given.

Error correction is a process which takes a state such as
$E_i \ket{\phi}$ to $\ket{\phi}$. Correction of $X$ errors
takes $X_u Z_v \ket{\phi}$ to $Z_v \ket{\phi}$; correction
of $Z$ errors takes $X_u Z_v \ket{\phi}$ to $X_v \ket{\phi}$.
Putting all this together, we discover the highly significant
fact that to correct {\em the most general possible} noise
(eq. (\ref{noise})), it is sufficient to
correct just $X$ and $Z$ errors. See eq. (\ref{synext}) and following
for a proof. It may look like we are only
correcting unitary `bit flip' and `phase flip' rotations,
but this is a false impression: actually we will correct
everything, including non-unitary relaxation processes!

\section{Correction of general errors}

A general quantum error correcting code will be an orthornormal
set of $n$-qubit states (quantum codewords) which allow correction of all
members of a set $S = \{ E_i \}$ of {\em correctable errors}.
We are usually interested in the case where the correctable errors
include all errors (including $X$, $Y$ or $Z$ and combinations thereof
for different qubits) of weight up
to some maximum $w$. Such a code is called a $w$-error correcting
code. It can be shown that such a code allows perfect correction
of any change in the $n$-qubit quantum state which affects at most
$w$ qubits \cite{96:SteaneB,96:Calderbank}. The correction method is
comparable to that shown in
figure 1 for the simple 3-bit code, where first a syndrome is extracted, and
then the state is corrected accordingly.

The main steps in the proof of this claim are as follows. Let
$\ket{\phi}_L$ be a state consisting of a general superposition of
codewords of some QEC code. A general noise process (\ref{noise})
will then produce the state \beq \sum_i \left(E_i \ket{\phi}_L
\right) \ket{\psi_i}_e  \label{synext} \eeq The syndrome
extraction can be done most simply by attaching an $n-k$ qubit
ancilla $a$ to the system (this is not the only method, but is
convenient), and storing in it the parity checks which are
satisfied by the quantum codewords, using controlled-not and
Hadamard operations. Another way of saying the same thing is that
we store into the ancilla the eigenvalues of a set of
simultaneously commuting operators (the {\em stabilizer}) acting
on the noisy state. The quantum codewords themselves are all
simultaneous eigenstates with eigenvalue 1 of all the operators in
the stabilizer, so such a process does not have any effect on a
general noise-free state. In the case of a noisy state, the effect
is to couple system and environment with the ancilla as follows:
  \beq \ket{0}_a \sum_i \left(E_i \ket{\phi}_L \right)
\ket{\psi_i}_e \rightarrow \sum_i \ket{s_i}_a \left(E_i
\ket{\phi}_L \right) \ket{\psi_i}_e.
  \eeq
The $s_i$ are $(n-k)$-bit binary strings, the syndromes. A
projective measurement of the ancilla will collapse the sum to a
single syndrome taken at random: $\ket{s_i}_a \left(E_i
\ket{\phi}_L \right) \ket{\psi_i}_e$, and will yield $s_i$ as the
measurement result. To keep the argument clear, we have assumed
that every $E_i$ will have a different syndrome. This assumption
will be dropped in a moment. Since there is only one $E_i$ with
the syndrome we have measured, we can deduce the operator
$E_i^{-1}$ which should now be applied to correct the error. The
resulting state is $\ket{s_i}_a \ket{\phi}_L \ket{\psi_i}_e$, and
now we can ignore the state of environment and ancilla without
affecting the system, since they are in a product state. The
system is therefore ``magically" disentangled from its
environment, and perfectly restored to $\ket{\phi}_L$!

This remarkable process can be understood as first forcing
the general noisy state to `choose' among a discrete set of
errors, via a projective measurement, and then reversing
the particular discrete error `chosen' using the fact that
the measurement result tells us which one it was.
Alternatively, the correction can
be accomplished by a unitary evolution consisting of
controlled gates with ancilla as control, system as target,
effectively transferring the noise (including entanglement
with the environment) from system to ancilla.

In general the true situation is that more than one error operator
has the same syndrome, just as in equation (\ref{sstate}). If two
error operators $E_1$, $E_2$ have the same syndrome $s$ for a
given QEC code, then upon obtaining $s$ in the syndrome extraction
we apply to the qubits the inverse of whichever of $E_1$ and $E_2$ was most
likely to have been produced by the noise process. The other error
operator is then an uncorrectable error. This will be
considered further in section \ref{s:success}.

\subsec{5.1 Example codes}

The above abstract treatment is clarified by some specific
examples. In section 3 we considered a simple example where {\em
either} an $X$ error {\em or} a $Z$ error (on any single qubit)
was correctable. The most important generalisation is to the case
of a code which allows {\em both} an $X$ error {\em and} a $Z$
error to be corrected. We will then further generalise to the case
where errors affecting more than one qubit are correctable.

We have already noted that $HZH = X$ and this means that if the
set of states (quantum error correcting code) $\{ \ket{u} \}$ are
to be correctable for $X$ and $Z$ errors, it is sufficient and
necessary that $\{\ket{u}\}$ and $\{ {\bf H} \ket{u} \}$ both be
correctable for $X$ errors, where ${\bf H} = H \otimes H \otimes H
\cdots \otimes H$ signifies the Hadamard rotation applied to all
$n$ qubits. This was how the first general class of QEC codes was
discovered \cite{96:SteaneB,96:Calderbank}. A general formalism in
which such ideas can be handled succinctly has been developed by
Gottesman \cite{96:Gottesman} and Calderbank {\em et al.}
\cite{98:Calderbank}. The simplest (though not the shortest)
single-error correcting code is the 7-bit code having two
codewords given by
\begin{eqnarray}
\left| {0} \right>_L  &=&
\left| {0000000} \right> + \left| {1010101} \right>
 + \left| {0110011} \right> + \left| {1100110} \right> \nonumber \\
&& \!\!\!\!\!\! + \left| {0001111} \right>  + \left| {1011010} \right>
+ \left| {0111100} \right> + \left| {1101001} \right>, \nonumber   \\
\left| {1} \right>_L &=& X_{1111111} \left| {0} \right>_L .
\label{7bit}
\end{eqnarray}
This code is constructed by combining two classical error
correcting codes: the states occurring in $\ket{0}_L$ are all the
members of a classical error correcting code $C$ which has minimum
distance 4, and the states occurring in ${\bf H} \ket{0}_L$ are all the
members of another classical error correcting code $C^{\perp}$
(the dual of $C$) which has minimum distance 3 and which contains
$C$. Furthermore, any
superposition of $\ket{0}_L$ and $\ket{1}_L$ only involves members
of $C^{\perp}$, and this is also true for any superposition of
${\bf H} \ket{0}_L$ and ${\bf H} \ket{1}_L$. It is not my purpose to present
this code construction fully, this may be found in the
references given in the introductory sections above. The aim is to give
enough information here to enable an understanding of the main
themes to be gained.

I referred just now to the concept of `minimum distance'. For a
set of binary words, the minimum distance is the minimum number of bits
which must be flipped to convert one binary word in the set
into another in the set. A classical error correcting code of distance $d$
can correct any set of bit flips as long as they affect no more than
$(d-1)/2$ bits. In the quantum case it can be shown that codes can
be constructed which allow correction of any error operator $E$ of
weight less than or equal to $(d-1)/2$, where $d$ is now the
minimum distance of the quantum code.

The code given by equation (\ref{7bit}) has distance $3$ and is
therefore a single-error-correcting quantum code. Since there are
two orthogonal states in the code, only a single qubit of
information can be stored in the 7 qubits employed. These
parameters are summarised by the notation $[[n,k,d]] = [[7,1,3]]$.
The code construction methods are now quite advanced. Infinite
sets of codes of given distance are known, as well as a large set
of quite powerful codes \cite{98:Calderbank,99:SteaneC}. Some
parameters of known codes include, for example, $[[5,1,3]]$,
$[[8,3,3]]$, $[[21,6,5]]$, $[[23,1,7]]$, $[[127,29,15]]$,
$[[127,43,13]]$. The 5-bit $[[5,1,3]]$ code has the minimum number
of bits needed to allow correction of a single general error.

\section{Success probability} \label{s:success}

As I have remarked in the introduction, the properties of QEC
codes, such as the set of correctable errors and the syndrome
extraction, follow immediately from the basic principles of
quantum mechanics. If a channel truly produces noise of the form
(\ref{noise}) where all the error operators are in the correctable
set for some code, then that code will certainly allow error-free
communication down the channel. However, any realistic channel
will produce not only correctable errors, but also uncorrectable
ones. The essence of evaluating the success of QEC is to calculate
the probability that the error correction will work, and this is
identical to calculating the probability $P_u$ that the channel
will produce an uncorrectable error. The fidelity of the final
state of the system of qubits, after passing through the channel
and having correction applied, will be $1-P_u$. For the rather
artificial channel considered in section \ref{s:3bit}, $P_u = 3p^2
- 2p^3$, and for the more realistic case of random rotations
through angles $\epsilon \phi$, considered in section 3.1, we
obtained $P_u \simeq 3p^2 = (2 \pi \epsilon)^4 / 3$ in the limit
of small $\epsilon$.

More general types of noise can be modeled in a number of ways.
As remarked before, by `noise' we mean simply any unknown or
unwanted change in the density matrix of our system. One approach
to modeling it is to consider the Hamiltonian describing a
coupling between the set of qubits and their environment. For
example, if the qubits are two-level atoms, we might consider the
effects of spontaneous emission of photons, adopting a master
equation approach.

The statement (\ref{noise}) about digitization of noise is
equivalent to the statement that any interaction between a system
of qubits and its environment can be written in the form
  \beq H_I =
\sum_{i} E_i \otimes H_i^e  \label{HI} \eeq
  where the operators
$H_i^e$ act on the environment. Under the action of this coupling,
the density matrix of the system (after tracing over the
environment) evolves from $\rho_0$ to $\sum_i a_i E_i \rho_0
E_i^{\dagger}$. QEC returns all terms of this sum having
correctable $E_i$ to $\rho_0$. Therefore, the fidelity of the
corrected state, compared to the noise-free state $\rho_0$, is
determined by the sum of all coefficients $a_i$ associated with
uncorrectable errors.

For a mathematically thorough
analysis of this problem, see \cite{96:Knill,97:Knill,00:KnillB}. The
essential ideas are as follows. Noise is typically a continuous process
affecting all qubits all the time. However, when we
discuss QEC, we can always adopt the model that the syndrome
is extracted by a projective measurement.
Any statement such as `the probability that error $E_i$ occurs' is
just a short-hand for `the probability that the syndrome extraction
projects the state onto one which differs from the noise-free
state by error operator $E_i$'. We would like to calculate
such probabilities.

To do so, it is useful to divide up (\ref{HI}) into a sum
of terms having error operators of different weight:
\beq
H_I = \sum_{{\rm wt}(E)=1} E \otimes H_E^e
+ \sum_{{\rm wt}(E)=2} E \otimes H_E^e
+ \sum_{{\rm wt}(E)=3} E \otimes H_E^e + \ldots   \label{HIwt}
\eeq
There are $3n$ terms in the first sum, $3^2 n!/(2! (n-2)!)$ terms
in the second, and so on. The strength of the system-environment
coupling is expressed by coupling constants which appear
in the $H_E^e$ operators. In the case that only the weight 1 terms
are present, we say the environment acts independently on the qubits:
it does not directly produce correlated errors across two or more
qubits. This language has lead to some confusion, since clearly
the environment is acting on all the qubits at once, and sometimes
the term `spatially correlated' is used to refer to any
situation where one degree of freedom (such as a mode of the
environment) can interact with all the qubits.
Nevertheless, if the interaction only contains error operators of
weight 1, then in the terminology of QEC we have a situation of
independent or uncorrelated errors. The important point is that
in this case, although errors of all weights will still appear
in the density matrix of the noisy system,
the size of the terms corresponding to errors of weight $w$ will
scale as $O(\epsilon^{2w})$, where $\epsilon$ is a parameter giving
the system-environment coupling strength.

As an example, consider dephasing produced by coupling
between a set of qubits and a bath of oscillators:
\[
H_I (t) = \hbar \sum_{q=1}^n \sum_k \sigma_{z,q} \left(
g_{kq}^* e^{-i \omega_{kq}t}a_{kq}
+ g_{kq} e^{i \omega_{kq}t} a_{kq}^{\dagger} \right)
\]
where $g_{kq}$ are the coupling coefficients, $a_{kq}$ are the
lowering operators for the modes. Each mode here couples to all
the qubits, but the error operators $\sigma_z$ in the Hamiltonian
are all of weight 1, and the resulting system density matrix will
be of the form $\sum_i a_i E_i \rho_0 E_i^{\dagger}$ where for
small times $t$,
\[
a_i \sim O \left( \left| \sum_{q=1}^n \sum_k g_{kq} \right|^{2
\,{\rm wt}(E_i)} \right).
\]

Since QEC restores all terms in the density matrix
whose errors are of weight $\le t=(d-1)/2$, the
overall failure probability $P_u$ is
the probability $P(t+1)$ for the noise to generate
an error of weight $t+1$ or more.
In the case of uncorrelated noise, this probability is approximately
\beq
P(t+1) \simeq \left( 3^{t+1}
\left( \!\! \begin{array}{c} n \\ t+1 \end{array} \!\!\right)
\epsilon^{t+1} \right)^2    \label{penv1}
\eeq
when all the single-qubit error amplitudes can add coherently
(i.e. the qubits share a common environment), or
\beq
P(t+1) \simeq 3^{t+1}
\left( \!\!\begin{array}{c} n \\ t+1 \end{array} \!\!\right)
\epsilon^{2(t+1)}         \label{penv2}
\eeq
when the errors add incoherently (i.e. either separate environments,
or a common environment with couplings of randomly
changing phase). The significance of (\ref{penv1}) and
(\ref{penv2}) is that they imply QEC works extremely
well when $t$ is large and $\epsilon^2 < t/3n$. Since good
codes exist, $t$ can in fact tend to infinity while
$t/n$ and $k/n$ remain fixed. Therefore as long as the noise
per qubit is below a threshold around $t/3n$, almost
perfect recovery of the state is possible. The ratio
$t/n$ constrains the rate $k/n$ of the codes which can
be constructed.

As a numerical example, consider the $[[127,29,15]]$ code. Let us
suppose we employ this code to store a general $29$-qubit state in
a collection of 127 two-level atoms, and we envisage that the excited state
of each atom is subject to spontaneous decay with a lifetime of 1
second. If the sequence of operations required for
syndrome extraction can be completed in 1 millisecond, then the
error probability per atom is of order $p=10^{-3}$. The
probability that an uncorrectable error occurs is approximately
$(3 n p)^8 /8! \simeq 10^{-8}$. Therefore we could repeatedly
correct the state for $10^5$ s$=1$ day before the fidelity of the
stored state falls to a half, despite the fact that during such a
process every one of the atoms will have decayed on average
about 50,000 times!

Such uncorrelated noise is a reasonable approximation in many
physical situations, but we need to be careful about the degree of
approximation, since we are concerned with very small terms of
order $\epsilon^d$. If we relax the approximation of completely
uncorrelated noise, equations (\ref{penv1}) and (\ref{penv2})
remain approximately unchanged, if the coupling constants in
(\ref{HIwt}) for errors of weight $t$ are themselves of order
$\epsilon^t / t!$ \cite{98:Aharonov}.

A very different case in which QEC is also
highly successful is when a set of
correlated errors, also called burst errors,
dominate the system-environment coupling, but we can find
a QEC whose stabilizer includes all these correlated errors.
This is sometimes called `error avoiding' rather than `error
correction' since by using such a code, we don't even need
to correct the logical state: it is already decoupled from
the environment. The general lesson is that the more
we know about the environment, and the more structure there
exists in the system-environment coupling, the better able
we are to find good codes.

The obvious approach to take in practice is a combined one,
in which we first discover the correlated contributions to the noise in our
system, and design a first layer of encoding accordingly, and then
overlay a second layer optimized for minimum-distance coding.
Such ideas have been studied in classical coding theory, where sometimes many
layers of encoding are combined, including tricks such as not
placing adjacent physical bits in the same logical block when the
code is not designed for burst errors.

The process of
encoding one bit in several, and then encoding each of those bits,
and so on, is referred to as code {\em concatenation}. When used
recursively, we obtain a powerful code whose behaviour
is relatively simple to analyse. This is the structure which
underlies the ``threshold result'', which states that arbitrarily
long quantum computations can be made reliable by introducing
more and more layers of concatenation, conditioned only
that the level of
noise per time step and per elementary operation on the physical
hardware is below a finite threshold. In other words, we
don't require a more and more precise and noiseless computer
if we want to evolve longer and longer computations. However,
we do need a bigger one.

\section{Fault tolerant methods}

There is ongoing research on constructing quantum error correcting
codes, and understanding the bounds on their size. In particular,
there are interesting cases of codes or correction methods which are not
covered by the stabilizer formalism.

In another direction, we are interested in codes whose
construction is thoroughly understood, but we wish to consider
what happens when we take a further step towards a realistic model
of a quantum computer. This further step is to allow not only the
channel, but also all the quantum gate operations involved in the
corrective process to be themselves noisy. It was far from obvious
whether QEC could still succeed in such conditions, since if we
`correct' a set of qubits on the basis of a faulty syndrome, we
will in fact introduce more errors.

Methods which can have a good probability of success even when all
the operations involved are imperfect are called `fault tolerant'.
Fault tolerance is possible through the use of simple ideas such
as repetition, and also some more subtle ideas which allow us to
construct networks of operations in which the routes by which
errors can propagate around a set of qubits are restricted. The
main themes were proposed by Shor \cite{96:Shor} and helpfully
summarised by Preskill \cite{98:Preskill,Bk:Lo}. A significant
collection of further ideas have been put forward by Gottesman
\cite{98:GottesmanA,99:GottesmanB}, which allow fault tolerant
methods to be found for a wide class of QEC codes, and the methods
were further improved in \cite{97:SteaneA,99:SteaneB}.

It was shown in \cite{99:SteaneB} that a quantum computer whose
size was a factor 22 larger than the number of logical qubits
required for its computations could be stabilized sufficiently to
allow large computations once the error probability per elementary
gate operation was of order $10^{-5}$, and the error probability
for every other qubit not participating in the gate was of order
$10^{-7}$ during the time of one gate. The analysis was based on a
$[[127,29,15]]$ code. To calculate the overall failure probability
it is necessary to track errors through networks of many hundreds
of elementary gates, and it is only possible to do this
analytically in an approximate way. I have now written a
(classical) computer code to automate such error tracking
calculations, and this has allowed the analytical approximate
methods to be checked for the case of smaller codes such as
$[[23,1,7]]$. This gives some confidence that the estimates in
\cite{99:SteaneB} are correct for the codes and networks
considered there.

\begin{figure}[ht]
\centerline{\resizebox{!}{7 cm}{\includegraphics{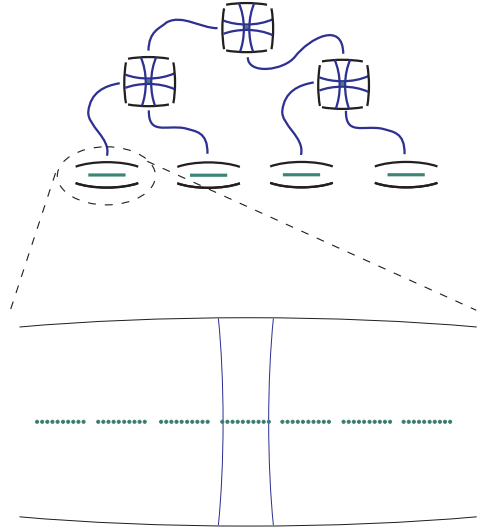}}}
%\rule{0pt}{24 pt}
\caption{Hierarchical design for a quantum computer. The qubits
are in groups of about 10. The physical mechanism of the computer
allows gates between any pair in a group, and between the end
qubits of each group and their nearest neighbours. Around 10 such
groups form a block (the illustration shows 7 groups per block).
Each block is coupled to the others by another physical mechanism,
such as an optical method, and this allows quantum communication
across long distances in the computer. To allow any one block to
communicate with one of several others, switching networks may be
necessary, as illustrated.}
\end{figure}

However, any such estimates involve
assumptions not only about what codes and fault tolerant
techniques are employed, but also about what degree of
communication and of parallelism
the physical hardware of the quantum computer allows. By
communication we mean the ease with which one qubit can be coupled
in a quantum gate with any other, not necessarily a neighbouring
one. By parallelism we mean the degree to which operations can be
carried out simultaneously on different sets of qubits. The case
considered in \cite{99:SteaneB} assumed that controlled multiple-not
operations, in which a single control qubit is coupled to multiple
target qubits, would be possible within each block of 127 qubits
in a single time step,
that operations in different blocks could be done in parallel, and
that qubits in one block could be coupled to qubits in any one of
several other blocks without introducing extra operations to cover
communication problems. Unfortunately, it is not clear whether
these assumptions are realistic.

High parallelism and ready communication tend to be mutually
exclusive. At one extreme is the case of only nearest-neighbour
interactions, i.e. limited communication, but any number of
(logically mutually consistent) operations can be carried out in
parallel. It was shown in \cite{00:Gottesman} that in this case
fault-tolerant methods can succeed, but no estimate of the noise
threshold has yet been made. It is certainly not as high as the
figures just quoted, which assumed much better communication. At
the other extreme is the case of perfect communication, that is
any qubit can communicate directly with any other, through a
single shared degree of freedom such as a photon mode or the
motion of ions in a single trap, but with no parallelism for gates
involving more than one qubit. It is unlikely that this case can
be stabilized sufficiently to allow large quantum computations.

Obviously the most robust physical system will be one which
combines good parallelism with good communication, but the extreme
case of full parallelism and direct communication between every
pair of qubits is almost certainly not optimal because any process
which allows such a high degree of inter-qubit coupling will have
correspondingly high unwanted coupling to noisy degrees of
freedom. I propose that the best construction for a quantum
computer is a {\em hierarchical} one, as illustrated in figure 2.
The most important role of the parallelism and communication
inside the computer is to get the syndromes out as quickly as
possible, so that qubits are not left accumulating noise for too
long before they are corrected. The syndrome extraction operation,
when carried out in a fault-tolerant manner, is itself
hierarchical, because many two-bit logic gates are needed between
qubits in each ancilla block, while coupling between one block and
another is needed less frequently, and distant blocks need to
communicate the least. The idea of the hierarchical quantum
computer is that the physical hardware should reflect this logical
state of affairs if an optimal robustness against noise is to be
achieved.

{\vspace{12pt}\noindent {\bf Acknowledgements}\vspace{12pt}}

I thank the organisers of the workshop. This work was supported by
EPSRC and the European Community network QUBITS.

\bibliographystyle{unsrt}
\bibliography{quinforefs}

\end{document}